# Unboxing Trustworthiness through Quantum Internet

**Agustín Zaballos, Adrià Mallorquí and Joan Navarro**

Engineering Department, La Salle Campus, Universitat Ramon Llull (URL), Barcelona, Spain

Research Group in Internet Technologies (GRITS)

{agustin.zaballos; adria.mallorqui; jnavarro}@salle.url.edu

ORCID: (0000-0003-2755-4428), (0000-0002-8930-6211), (0000-0003-3916-9279)

**Abstract:** The broad adoption of the Internet of Things during the last decade has widened the application horizons of distributed sensor networks, ranging from smart home appliances to automation, including remote sensing or telemetry. Typically, these distributed systems are composed of several nodes attached to sensing/acting devices linked by a heterogeneous communication network. The unreliable nature of these systems (e.g., devices might run out of energy or communications might become unavailable due to an insufficient fading margin) drives practitioners to implement heavyweight fault tolerance mechanisms to identify those untrustworthy nodes that are misbehaving erratically and, thus, ensure that the sensed data from the IoT domain are correct. The overhead in the communication network degrades the overall system, especially in scenarios with limited available bandwidth that are exposed to severely harsh conditions.

Quantum Internet might be a promising alternative to minimize traffic congestion and avoid worsening reliability due to the link saturation effect by using a quantum consensus layer. In this regard, the purpose of this paper is to explore and simulate the usage of quantum consensus architecture in one of the most challenging natural environments in the world where researchers need a responsive sensor network: the remote sensing of permafrost in Antarctica. More specifically, this paper 1) describes the use case of permafrost remote sensing in Antarctica, 2) proposes the usage of a quantum consensus management plane to reduce the traffic overhead associated with fault tolerance protocols, and 3) discusses, by means of simulation, possible improvements to increase the trustworthiness of a holistic telemetry system by exploiting the complexity reduction offered by the quantum parallelism. Collected insights from this research can be generalized to current and forthcoming IoT environments.

**Keywords:** Quantum Internet, Antarctica, Reliable data network.

## 1. Motivation of the research project

Antarctica is a key location for many research studies in several fundamental fields, such as oceanography, bioscience, geoscience, and other environmental studies domains that mainly require experimentation in the field. Although many bases have been settled in the Antarctic continent, its harsh environmental conditions suppose a big challenge when implementing and maintaining new operational services. Actually, it must be considered that Antarctic campaigns, aimed at carrying out scientific experiments in such a desolate environment, are usually restricted to the summer period due to adverse meteorological conditions. There, as in many other harsh environments, physical human interactions required for network troubleshooting (e.g., to fix architectural communication disarrangements) at usually inaccessible locations are often unfeasible or unaffordable due to the severe conditions this area is exposed to. This fact typically results in an unprecedented loss of critical data, experiments delay, or expensive fix-and-resolve campaigns.

The lack of existing telecommunication systems in this area to interconnect remote base camps hardens the possibility of building synergies among different polar experiments and limits the ability of researchers to evolve potential interdisciplinary research works. Besides, in Antarctic latitudes, the degree of coverage offered by existing communication solutions like Very Small Aperture Terminal (VSAT) satellite systems is far from optimal. This weakness greatly limits the feasibility of deploying a large-scale telemetry service to support several experiments, such as geomagnetic studies, climate change, biological monitoring, or permafrost analysis. Therefore, most of the measured data derived from these experiments are currently manually collected (i.e., a practitioner physically goes to the remote sensor location to download data), which obviously causes many logistic difficulties and limits the scope of the experiments. This situation has motivated new research approaches for automated and autonomous data gathering systems, mainly driven by ad hoc Wireless Sensor Networks (WSNs), to operate in such remote and hash environments where scalability and fault tolerance are paramount needs.

One of the well-known initiatives to address this challenge is the SHETLAND-NET research project [1], which aims to use a backhaul technology, called Near Vertical Incidence Skywave (NVIS) that expands the use of communications in HF (3-30 MHz) by ionosphere reflection [2] to establish a communication link between different Internet of Things (IoT) domains in the archipelago of the South Shetland Islands. Each of these domains is a Ground Terrestrial Network-Permafrost (GTN-P) station that measures the permafrost status (i.e., telemetry) in the Antarctic soil and publishes its local measurements through the NVIS link. It is worth noting that NVIS does not require direct vision because the information signal is transmitted upwards to the ionosphere. Thus, it allows for overcoming any geographical barrier, which results in a very convenient alternative when designing large-scale networks for infrastructure-less environments in terms of communications. For this reason, NVIS has a coverage range of up to 250 km and can easily achieve bit rates of up to a few thousand bps [2]. However, the reliability of NVIS is strongly dependent on ionosphere conditions and solar activity, which adds an extra level of complexity to the inherent harsh conditions of sensing and processing data in the Antarctica environment. This results in an unprecedented (in the field of communication networks) number of end-to-end disconnections, exceptionally variable error rates, and highly intermittent connectivity.

The combination of these drawbacks, together with the inherent extreme conditions of the environment, originate low levels of network performance and trustworthiness over traditional TCP/IP architectures used in current communication networks. Additionally, these effects can vary and become considerably worse depending on the exact physical location where the experiment—that will generate telemetry data to be processed in the control center—is deployed in the Antarctic soil. Hence, anticipating the possible issues and difficulties related to communication before deploying the Antarctic experiments, that is, assessing the expected trustworthiness of the data communication architecture, has become a hot research challenge. Indeed, Antarctic campaigns are usually very time-restricted due to meteorological conditions, and possible arisen difficulties are very challenging to overcome. Thus, it is necessary to study the viability and the expected trustworthiness of implementing this kind of network before its deployment in the field [1][2].

This paper explores the use of quantum technologies to improve the overall performance and trustworthiness of the communication protocol architecture and enable technologies required to implement the Antarctic telemetry service reliably. More specifically, the proposed architecture enables to update the degree of trustworthiness of the different elements that articulate the system (e.g., communication links, concentrator nodes, sensing devices) by 1) proposing how the quantum-related network architecture could drive the maximization of IoT trustworthiness and 2) processing the information measured from different spots and, accordingly, reducing the amount of data transmitted from non-trustable regions. Riverbed Modeler network simulator [3] has been used to assess, in advance, the success of our experimentation in the field under the specific Antarctic conditions collected from SHETLAND-NET when implementing the remote sensing of permafrost.

Honestly, our research has pushed us to explore options beyond what is strictly expected to be deployed in the short term and evaluate all the possibilities that the quantum scenarios offer to this simulated environment. The drawbacks of the harsh Antarctic Internet and the extreme working conditions challenge us to evolve our previous trustworthiness model and take profit from quantum Internet's main advantages to overcome low-performance levels. The main contribution of this paper is the proposal of a definition of quantum Internet architecture to support an IoT telemetry service deployed in an Antarctic wireless sensor network and the evaluation of quantum Internet paradigm benefits in the trustworthiness model. The existing architecture deployed by the SHETLAND-NET project—that must work hand in hand with the proposed quantum Internet one—suffers from some properties similar to those of well-known heterogeneous networks. In this sense, a layer-based quantum trustworthiness model is helpful for assessing and improving the expected research outcomes to address this challenge. To sum up, this quantum-related research pursues the overarching challenge of maximizing the trustworthiness of Antarctic IoT environments, which could be further generalized to Internet communications.

The remainder of this paper is organized as follows. Section 2 describes the quantum Internet protocol architecture necessary to alleviate the overhead associated with fault tolerance. Section 3 proposes a model for trustworthiness using quantum technologies. Section 4 proposes a set of key performance indicators to assess the proposed trustworthiness model over the quantum internet by means of simulations. Finally, Section 5 evaluates the whole proposal and discusses the major findings and shortcomings.

**2. Quantum Internet protocol architecture**

Permafrost is a good indicator of global climate change and overall Antarctic warming. However, its large-scale measurement is still challenging due to the difficulties derived from reading data through the currently deployed Antarctic Internet [1][4]. The sensing technology used at GTN-P stations for permafrost monitoring being used nowadays requires a trustworthy data flow in order to enable a continuum and meaningful analysis [1][4], which greatly limits the affordable number of human interactions to fix transmission impairments or network troubleshooting [4]. Currently, this service is implemented by a set of distributed LoRa-based sensors that collect information regarding the status of the frozen ground by using an Antarctic Internet access network held up by a backbone infrastructure to acquire and provide permafrost information. This Internet backbone network is currently implemented by an NVIS link, which showed to be insufficient for continuous data transmissions due to its inherent unreliability [1][2][4]. This section (1) discusses the current limitations of existing communication technologies in the Antarctic for the permafrost telemetry use case (which can be extended to other forthcoming services) and (2) proposes a suitable quantum Internet protocol architecture to address these limitations.

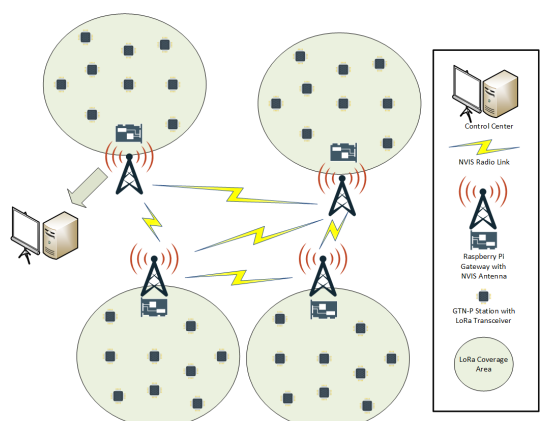

**Figure 1.** Global Wireless Sensor Network in the Antarctic experiment.

*2.1 Current limitations of communication technologies for permafrost telemetry in Antarctica*

Antarctic sensor networks must be designed to face and overcome the harsh conditions its telecommunication infrastructure will be exposed to. In the Antarctic experiment, all remote sensors need to be interconnected to a control center. Currently, this is done using a heterogeneous Global Wireless Sensor Network (GWSN). In this regard, The SHETLAND-NET project has implemented a proposal for an Antarctic Internet with a heterogeneous architecture that integrates WSN, Delay Tolerant Network (DTN), and NVIS. This combination enables low-rate data services (that might be suitable for telemetry) with a low-cost infrastructure that can be easily deployed. Hence, as shown in Figure 1, distributed sensor nodes interact with each other and with the backbone infrastructure to acquire, process, transfer, and provide information collected from the physical world. As some of these sensor nodes can also provide processing and concentrator capabilities—considering their limitations in terms of power, memory, and processing resources—when the NVIS communication backbone is unavailable, the usage of the DTN becomes very appropriate. In fact, the DTN running on top of the WSN and the NVIS link enables the system to opportunistically send through the Internet all the data collected when the ionosphere is unavailable for data transmission. Note that this process might result in additional network congestion and packet loss, which unavoidably degrades the overall trustworthiness of the data services running on top of this communication system [5][6].

In this regard, one of the key design parameters in this domain is trustworthiness, which can be best seen as the system's ability to provide reliable results. For the specific use case of permafrost telemetry, it has been acknowledged that to ensure continuum data monitoring, the probability of successful communications has to be over the 60% [1][4].

*2.2 Quantum Antarctic Internet proposed for permafrost telemetry*

This section proposes an effective quantum Internet scheme to enable automatic telemetry for multiple GTN-P measuring spots between the GWSN and the telemetry Control Center (Figure 2). Such quantum Internet will work in synergy with classical Internet to overcome the limitations of traditional Internet technologies. In this way, the communication model conceived for quantum teleportation relies on the tight integration of both classical and quantum operations and communications despite the entanglement mechanism not having any counterpart in the classical Internet. In fact, the quantum Internet can be best seen as an interconnection of quantum devices and repeaters that can generate and exchange quantum information in the form of quantum bits, called qubits. Thus, the quantum Internet is based on two important features of quantum entanglement [7][8]:

- It is inherently private by the laws of quantum mechanics (i.e., no-cloning theorem) and hence avoids eavesdropping attacks. Many experimental studies have demonstrated that long-distance private key sharing via quantum networks can succeed (e.g., quantum key distribution) [9][10].
- It provides instantaneous coordination of the communication parties (i.e., quantum teleportation). This feature is of utmost importance in our research for maximum reliability of data transmission by using the consensus mechanism [11]. It turns out that qubit teleportation requires distributing a pair of Bell particles (one to the source and the other to the destination). When they are entangled (i.e., each entangled qubit's state cannot be mathematically described independently), the collapse of the wave function of a qubit, due to a measurement action, immediately affects the other regardless of the physical distance between them (although this result is communicated to the destination using a classical channel so that the recipient could use it to decide what transformations it has to carry out to recuperate the original data qubit).

A protocol stack must be developed to structure end-to-end quantum communication issues and tackle complex requirements of quantum applications over quantum Internet. Figure 3 shows the proposed quantum Internet heterogeneous protocol architecture, which visually describes how functions are vertically composed to achieve quantum communication [12][13]. This quantum Internet architecture can provide new capabilities that are not reproducible by classical systems. Although the fundamental quantum Internet technology is not yet mature, research on quantum Internet could be conducted. Note that the heterogeneous protocol architecture proposed in Figure 3 allows researchers to build their own applications by delivering a set of telemetry tools and development platforms that provide a complete data life cycle without worrying about the underlying hardware and software implementing these services [14].

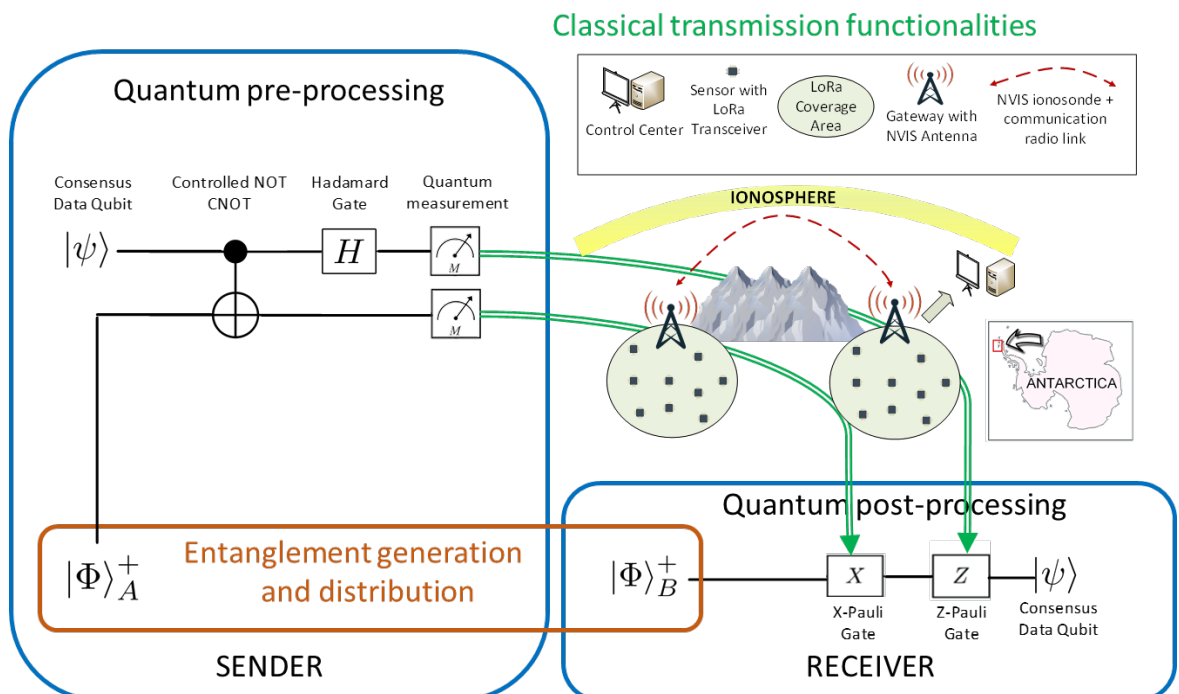

**Figure 2.** Conceptual scenario for quantum Internet deployment in Antarctica use case.

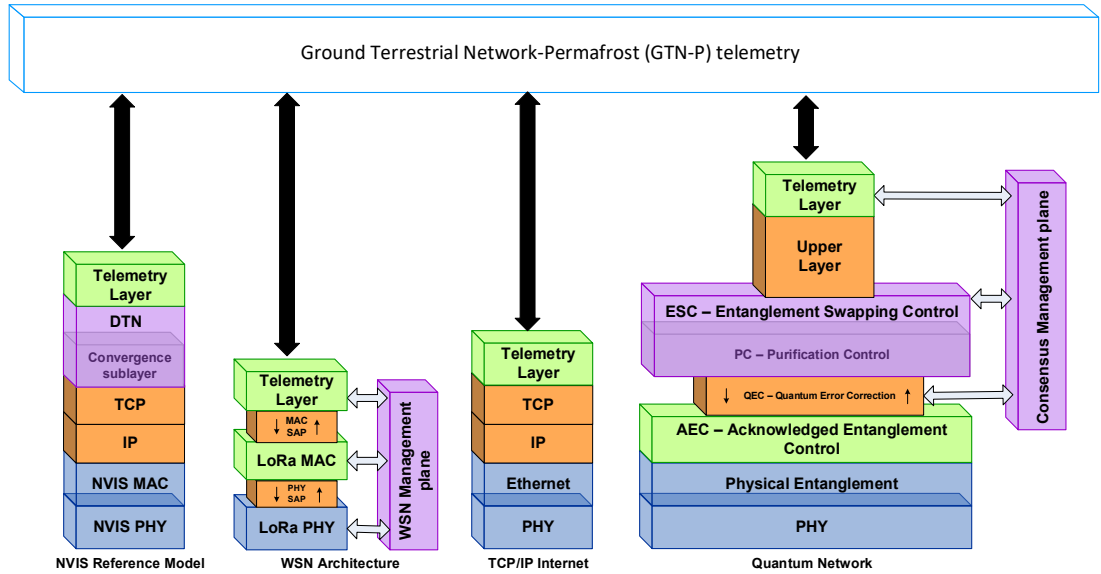

**Figure 3.** Quantum Internet protocol architecture in Antarctica for permafrost telemetry.

The permafrost telemetry service in Antarctica needs the long-distance quantum communication that entanglement-based physical technologies can carry out [15][16], even though the implementation of quantum Internet requires a novel design. The main goal of our work is to design and simulate a quantum Internet model to promote trustworthiness in a Measure Directly (MD) quantum scheme [17] in which there is no quantum memory to store entanglement, and qubits are immediately measured to produce classical correlations.

Physical and link layers are responsible for converting quantum information into flying qubits by employing different encoding methods. Entanglement between neighbor quantum nodes has to be established to generate adjacencies by using short-distance entanglements and an Acknowledgement Entanglement Control (AEC) that controls entanglement distribution as defined in [18]. It should be said that while the physical and link layers of quantum networks require experimental assessments on costly hardware, the generation and the distribution of entanglement can be assessed via simulations if the simulation environment is consistently modeled with the practical physical facts [19]. In this regard, simulations must take into account that the physical layer handles the physical entanglement (e.g., entanglement generation, timing synchronization, and heralded entanglement), and the link layer promotes robust entanglement generation. At the link layer, quantum error correction mechanisms and entanglement purification control tackle the attenuation of entanglement fidelity due to quantum decoherence effects. The entanglement network structure is completely independent of the underlying physical channel configuration [12].

The modeled network layer performs entanglement swapping that oversees binding multiple link-level entanglements together by using an Entanglement Swapping Control (ESC) [18]. Long-distance entanglement, adjacencies topology control, and entanglement swapping peering can be easily modeled using a QoS-aware reactive routing protocol paradigm [20]. The first task of a Quality of Service (QoS) routing protocol is to find several suitable loop-free paths from the source to the destination with the necessary available resources to meet the QoS requirements of the desired service with strict constraints of QoS. The constraint-based routing protocols find one or more paths that satisfy a subset of QoS conditions imposed by the user within a definition of a QoS constraints scope. They can be applied to a single link or the whole path [21].

Routing algebra and policies are required to define the routing protocol behavior formally. Our proposal is based on Sobrinho's routing algebra [22][23]. Note that an algebraic approach is very useful for both understanding existing protocols and for exploring the design space of future Internet routing protocols. The proposed routing policy is formed by:

$$A = <\Sigma, \oplus, L, \leqslant> \quad (1)$$

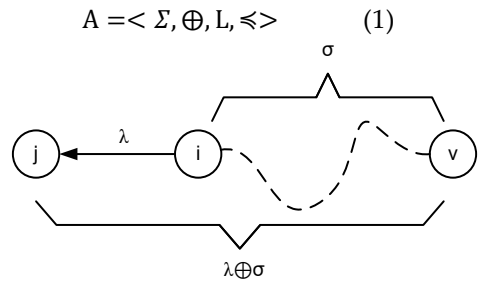

**Figure 4.** Example of the routing algebra where $(\lambda \in L) \wedge (\sigma \in \Sigma)$

**Table I**. Elements of the routing policy

| Element | Description |
|---|---|
| $\Sigma$ | It is the cost associated with a path, and it is known as the signature. |
| $\oplus$ | It defines the way to add the cost of a link to a path to calculate the total cost. It is known as the operator. |
| L | It represents the cost associated with a link, and it is known as the label of the link. |
| $\leqslant$ | It is the precedence relationship, and it is used to decide which path is the best one. |

Each element of this array (1) is defined in Table I. In addition to this, two logical operators are necessary: AND (∧) and OR (∨) operators. In this paper, the following model of a cost computation based on [23] is used. It is outlined in Figure 4, where *node j* is the destination of the routing information and *node v* is the origin.

Several routing protocols are able to provide QoS. All of them provide the QoS based on the metric used by the routing protocol, which gives more information than the number of hops. This is the reason why the selected quantum metric for QoS is an important parameter when a routing protocol is designed [24]. The metric is the value used to select which path is the best one. Moreover, it is necessary to define the optimization functions that define the routing policy's objective. An interesting quantum metric strategy is based on a scheme with multiple metrics that represents a link with more than one cost value [20][25]. The entire classical network exchanges this information, and then the nodes can combine this information in order to decide the best path to the destination. Table II proposes a multiple quantum metric example by using three main quantum metrics to build the routing policy: 1) the hop count to take into account the swaps chain and the number of deployed repeaters (and, consequently, the number of needed purifications that is highly dependent of the number of swaps considering that entanglement swapping would be essential in generating long-distance entanglement [26]), 2) the number of used Bell Pair per second for a certain fidelity that is highly dependent of construction materials and the corresponding decoherence effect), and 3) the average overhead introduced (i.e., added redundant qubits and extra needed quantum operations). A routing metric is considered optimal if a routing protocol exists that always discovers the most favorable path between any pair of nodes in any connected network when used in conjunction with such a metric.

**Table II**. Multiple quantum metrics scheme (lexicographic order)

| Σ | $\Sigma_{Hop} \times \Sigma_{BellPairs} \times \Sigma_{Overhead} : <\sigma_H, \sigma_{BP}, \sigma_O>$ |
|---|---|
| ⊕ | $(\lambda_{BP}, \lambda_O) \oplus (\sigma_H, \sigma_{BP}, \sigma_O) = <\sigma_H + 1, \lambda_{BP} + \sigma_{BP}, \lambda_O + \sigma_O>$ |
| L | $\lambda_{BP} \in R^+, \quad \lambda_{BP} = $ Bell Pair/seg <br> $\lambda_O \in R^+, \quad \lambda_O = $ total overhead |
| ≼ | $(\sigma_H, \sigma_{BP}, \sigma_O) \preccurlyeq (\sigma'_H, \sigma'_{BP}, \sigma'_O)$ iif <br> $(\sigma_H < \sigma'_H) \vee (\sigma_H = \sigma'_H \wedge \sigma_{BP} < \sigma'_{BP}) \vee$ <br> $(\sigma_H = \sigma'_H \wedge \sigma_{BP} = \sigma'_{BP} \wedge \sigma_O < \sigma'_O)$ |

In this way, the Antarctica permafrost network layer can finally be modeled as a minimal expression since our entanglement swapping uses predefined communication paths. So, issues such as monitoring the success rate of entanglement distribution and the implementation of the entanglement swapping, and the quantum fidelity-awareness routing (highly affected by the distance of the quantum channel) are out of the scope of our network layer model.

The modeled transport layer (see Figure 3) is focused on facilitating a deterministic transmission by implementing the Entanglement Swapping Control (ESC) and the Purification Control (PC) in order to tolerate fluctuating delays in entanglement generation, as defined in [18], and it is in charge of the congestion control. Congestion control aims to detect the network status before it suffers from an overwhelmed situation. There are two main strategies to detect and mitigate network congestion [5]:

- Reactive, typically used in connectionless packet-switched networks, in which no resource reservation is made prior to data transfer, and techniques are used to resolve the congestion once it is detected.
- Preventive, which is generally used in circuit-switched networks. Resources are reserved during connection setup to prevent congestion during data transfer, limiting the number of users and monitoring the flow so that it does not exceed a predetermined limit. This is the chosen option for our telemetry quantum Internet because of its simplicity and because an exhaustive simulation of the transport layer does not offer any remarkable contribution to our research further than the obtained in [5]. The network status estimation process enables to determine the connection's potential bandwidth, specifying the maximum bandwidth that will be available. It is calculated at the beginning of the transmission and provides a rapid setup of the sending rate, thus creating an immediate benefit in the throughput usage and optimizing its convergence. Trustworthiness classical affectations due to transport protocol are studied in [5].

## 3. Quantum trustworthiness modeling

*3.1 Trustworthiness model for classical Internet*

Internet trustworthiness is generally defined in the literature as the property of behaving as expected under adversarial conditions. In [27], five dimensions are introduced in the general trustworthiness concept:

1) The safety, that ensures that a system operates without causing unacceptable risk of damage to people's health, directly or indirectly, as the result of damage to property or the environment.
2) The security, that protects a system from unintended or unauthorized access, change, or destruction of data.
3) The reliability, that describes the system's ability to perform its required functions under stated conditions for a specified period.
4) The resilience, that describes the ability of a system to prevent any severe impact of a disruption while maintaining an acceptable level of service.
5) The privacy, that protects the individuals' right to control what information related to them may be collected.

The permafrost telemetry use case mainly focuses on the third dimension, reliability, and partially on the fourth dimension, resilience (see Figure 5). It is important to highlight that cyber-security issues regarding intentional malicious digital attacks are out of the scope of our trustworthiness analysis as these kinds of issues are beyond the definition of harsh environment used in this paper for the sake of Antarctic telemetry. The vulnerability assessment in our use case, when needed, is carried out by upper-layer protocols, and it is not just to confirm the security of a network and its devices; it also stems from the concern that a network might not be adequately protected from known threats. Understanding how threats and attacks can be performed against systems is necessary to design an appropriate security monitoring surveillance system [14]. Moreover, network security analysis must coordinate different sources of information to support effective security models. These sources of information should be distributed through the entire network in order to monitor and manage the maximum number of communications. Proper network protection demands performing periodic security tests to control devices and services and identifying possible vulnerabilities. Testing should be recurring throughout the campaign, and security experts should rely on solutions that provide thorough, flexible, quantifiable, repeatable, and consistent network security tests [14].

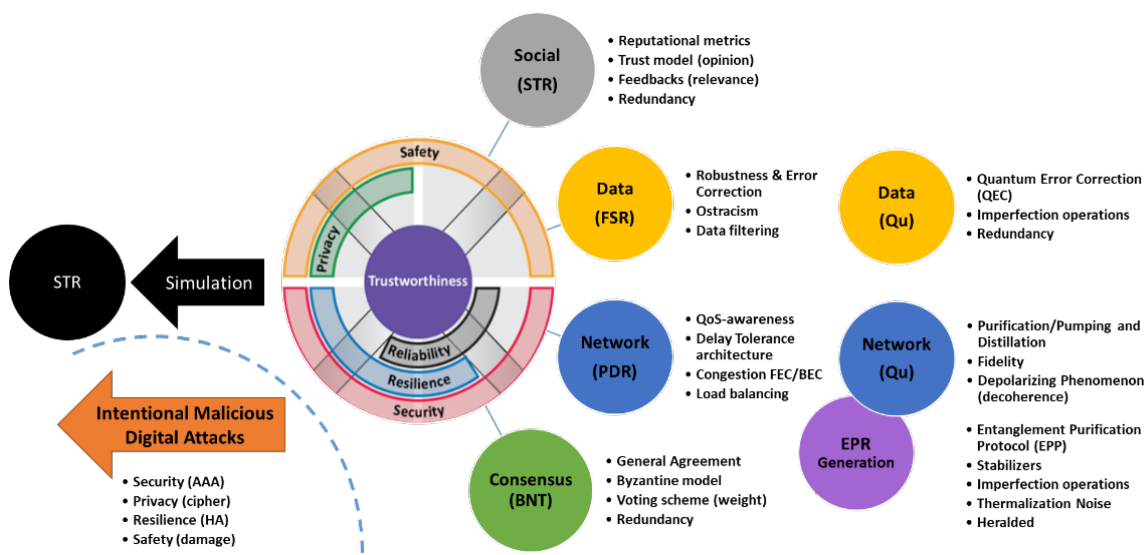

**Figure 5.** Trustworthiness dimensions and layers for classical and quantum Internet.

The proposed trustworthiness model that will be used to measure and evaluate the performance of the Antarctic quantum Internet is layer-based (Figure 5). This model comprises two baseline layers (data trustworthiness and network trustworthiness for the classical and quantum Internet) and two extension layers (social trustworthiness and consensus). Actually, data, network, social trustworthiness, and consensus layers are tributaries to the whole system's trustworthiness [11]:

1. Data trustworthiness layer: It is the layer responsible for ascertaining the correctness of the data provided by the source. Many methods in the literature try to detect faulty nodes, false alarms, and sensor misreading using different approaches [28].

2. Network trustworthiness layer: It is the layer responsible for assuring that a packet reaches its destination on time and unaltered despite the adversities (e.g., link failure or link saturation). The improvement of this layer is a challenge that has been addressed from different perspectives, such as transmission coding, load balancing and redundancy protocols, robust transport protocols, dynamic routing and topology control protocols, and DTN architectures [29][30][31].

3. Social trustworthiness layer: It is the layer responsible for leveraging the capability of the objects to establish social relationships autonomously between them to improve their trust according to the correctness of the collected data. This layer has gained more attention since the irruption of the Social Internet of Things (SIoT), in which the "things" have the capability to establish social relationships autonomously between themselves to define more complex trust and reputation models based on decentralized, self-enforcing trust management and reputational multiparty calculations [32][33].

4. Consensus Layer: It is the layer responsible for reaching a state where all the group participants agree on the same result. That is, a state where all participants of the same distributed system agree on the same data values through a general agreement despite classical byzantine faults. This layer implements voting-based mechanisms to reach that agreement. It is worth noting that one of the main drawbacks of these powerful mechanisms is the number of messages that need to be delivered throughout the network to reach an agreement [34][35].

To the best of our knowledge, previous trustworthiness approaches found in the literature focus on specific areas of trustworthiness, especially for the classical Internet. However, our previous studies [5][6][11] clear up the goodness of 1) the reputational algorithms used in the social trustworthiness layer, where complex trust models are used, and 2) the consensus algorithms, where all participants agree on data values through a general agreement. Although a combination of reputational and consensus algorithms offers the best trustworthiness performance in several situations, in the scenarios where traffic saturation exists, the consensus layer, which implements voting-based mechanisms, becomes one of the main drawbacks due to overload traffic. This situation drives us to explore the utilization of consensus algorithms without the detriment of traffic congestion that worsens performance and decreases Internet trustworthiness. This motivates the inclusion of quantum Internet in our trustworthiness model to explore the features such as instantaneous coordination of the communication parties for the sake of consensus algorithms.

*3.2 Trustworthiness model adaptation to the quantum Internet*

The main evolution of the proposed trustworthiness model is focused on the two baseline layers: data trustworthiness and network trustworthiness. The reliability at these baseline layers includes the processing of qubit loss (note that the reliable transmission of qubits could be addressed by direct forwarding and teleportation), error correction, and network congestion effects mitigation.

A. *Quantum data trustworthiness layer*

The smallest unit of quantum information is known as a quantum bit or qubit (Figure 6). Unlike classical computing, which is based on a bit and the system can be either a '*0*' or a '*1*', which switches only when an operation occurs, a single qubit can be in a superposition state of two states, simultaneously representing 0 and 1 state variables (e.g., the vertical polarization and the horizontal polarization of a photon or the spin-up and the spin-down of an electron). Measuring a qubit will destroy the superposition state and determine the state in a classical fashion. The measurement outcome of a quantum state is the eigenvalue of the eigenvector.

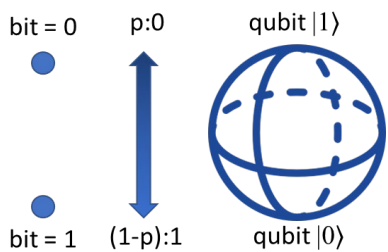

**Figure 6.** Bit: deterministic bits with two possible values (0 ∨ 1); Probabilistic bit: {p:0, (1-p):1}; Qubit: superposition of 0 and 1, $\alpha|0\rangle + \beta|1\rangle$, $\alpha, \beta \in \mathbb{C}$.

To understand one of the most important Di Vincenzo's criteria [36] – "well characterized qubit is needed" – some mathematical concepts about notation and quantum mechanics are needed. We must present the notation necessary for reading the protocols and proofs in this paper. The simplest quantum system is essentially a two-state system, and the single qubit pure state can be expressed using the Dirac notation. For a more detailed explanation of the relevant background, see [7] or a textbook such as [18].

In quantum mechanics, bra-ket notation is used to denote quantum states, typically represented as an element of a complex Hilbert space. In this way, linear functional ⟨φ| (bra) that acts on a vector |θ⟩ (ket) is written as ⟨φ||θ⟩∈ ℂ. The inner product, ⟨φ||θ⟩, is typically interpreted as the probability amplitude for the state θ to collapse into the state φ (i.e., the coefficient for the projection of θ onto φ). Actually, a ket is a vector in a Hermitian vector space (column vector), and a bra is a row vector where $|\varphi_A\rangle \in V$ and $\langle\varphi_A| \in V^\dagger$:

$$|\varphi_A\rangle = \begin{pmatrix} a_0 \\ \vdots \\ a_{N-1} \end{pmatrix} \qquad \langle\varphi_A| = (a_0^* \quad \ldots \quad a_{N-1}^*) = |\varphi_A\rangle^\dagger$$

Thus, for example, we can say that ⟨φ||φ⟩ = 1 and ⟨φ||θ⟩ = 0 if one orthonormal basis is used. Then,

$$\langle 1|0\rangle = 0 = \begin{pmatrix} 0 & 1 \end{pmatrix} \begin{pmatrix} 1 \\ 0 \end{pmatrix} = \sum_0^{N-1} a_i^* \cdot b_i$$

Where $a = x + jy = r * e^{j\varphi}$, $a^* = x - jy = r * e^{-j\varphi}$ and $|a| = r = \sqrt{x^2 + y^2} = \sqrt{a \cdot a^*}$.

Another interesting operation that will be used in the paper is the tensor product to form a third Hilbert space (for composite systems with more than one qubit). The tensor product is defined as:

$$|A\rangle \otimes |B\rangle = |AB\rangle = \begin{pmatrix} a_{0,0} & a_{0,1} \\ a_{1,0} & a_{1,1} \end{pmatrix} \otimes |B\rangle = \begin{pmatrix} a_{0,0} \cdot B & a_{0,1} \cdot B \\ a_{1,0} \cdot B & a_{1,1} \cdot B \end{pmatrix}$$

For example, if $|q_1\rangle = a|0\rangle + b|1\rangle$ and $|q_2\rangle = c|0\rangle + d|1\rangle$, then:

$$|q_1\rangle \otimes |q_2\rangle = ac|00\rangle + ad|01\rangle + bc|10\rangle + bd|11\rangle = \begin{pmatrix} ac \\ ad \\ bc \\ bd \end{pmatrix} = \begin{pmatrix} a \\ b \end{pmatrix} \otimes \begin{pmatrix} c \\ d \end{pmatrix}$$

Otherwise, if:

$$|q\rangle = \frac{\sqrt{3}}{2}|00\rangle + \frac{1}{2}|10\rangle = \begin{pmatrix} \sqrt{3}/2 \\ 0 \\ 1/2 \\ 0 \end{pmatrix} = \begin{pmatrix} \sqrt{3}/2 \\ 1/2 \end{pmatrix} \otimes \begin{pmatrix} 1 \\ 0 \end{pmatrix} = |q_1\rangle \otimes |q_2\rangle = |q_1\rangle \otimes |0\rangle$$

So, $q_1$ is not equiprobable, and $q_2$ is in the state $|0\rangle$, but they are independent and remain unaltered and not entangled. In fact, there exists only 4 two-qubit registered as entangled, the Einstein, Podolsky, Roser pair (EPR): $|\Psi_A^\pm\rangle$ and $|\Phi_A^\pm\rangle$ (2.a-d). When states are entangled, each qubit's state cannot be described independently. That is to say, the collapse of the wave function of a qubit by measurement may immediately affect the other of the pair's state regardless of the physical distance between them. Each qubit state has an equally weighted probability, so each qubit has a 50/50 probability of being found in each state but not independently. Therefore, measuring one qubit will also decide the other qubit's state.

$$|\Psi_A^+\rangle = \frac{|01\rangle + |10\rangle}{\sqrt{2}} = \frac{1}{\sqrt{2}}\begin{pmatrix} 0 \\ 1 \\ 1 \\ 0 \end{pmatrix} \quad (2.\text{a})$$

$$|\Psi_A^-\rangle = \frac{|01\rangle - |10\rangle}{\sqrt{2}} = \frac{1}{\sqrt{2}}\begin{pmatrix} 0 \\ 1 \\ -1 \\ 0 \end{pmatrix} \quad (2.\text{b})$$

$$|\Phi_A^+\rangle = \frac{|00\rangle + |11\rangle}{\sqrt{2}} = \frac{1}{\sqrt{2}}\begin{pmatrix} 1 \\ 0 \\ 0 \\ 1 \end{pmatrix} \quad (2.\text{c})$$

$$|\Phi_A^-\rangle = \frac{|00\rangle - |11\rangle}{\sqrt{2}} = \frac{1}{\sqrt{2}}\begin{pmatrix} 1 \\ 0 \\ 0 \\ -1 \end{pmatrix} \quad (2.\text{d})$$

A qubit is the simplest quantum mechanical system that can be represented geometrically by a Bloch sphere [7][18]. Any pure quantum state $|\psi\rangle$ can be represented by a point on the surface of the Bloch sphere with spherical coordinates of θ as a polar angle (i.e., the angle that a line makes with ẑ-axis), and Ø as an azimuthal angle (i.e., the angle that a line makes with x̂-axis). The range of values for θ and Ø such that they cover the whole sphere (without redundancies due to the symmetry of the sphere geometry) are θ∈[0,π) and Ø∈[0,2π). Angle θ corresponds to latitude, and angle Ø corresponds to longitude. Every individual state in the Bloch sphere is represented by a two-dimensional (2D) vector, where $|0\rangle$ and $|1\rangle$ are its basis vectors, which are orthogonal to each other (Figure 7). In fact, there are three mutually unbiased bases (i.e., three ways of encoding information in different ways, as is shown in Table III). Mixed quantum states are those states that provide probabilistic results when measured with all basis.

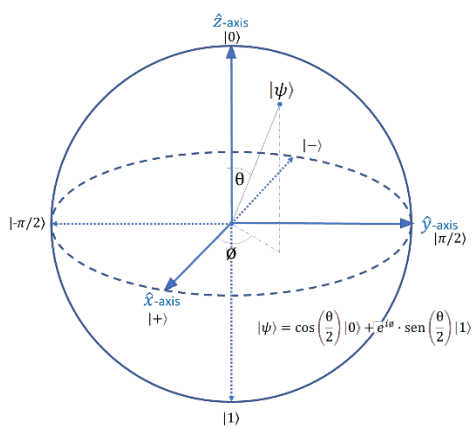

**Figure 7.** The Bloch sphere is a geometric representation of a single qubit pure state as a unit vector pointing on the surface of a unit sphere (3.a-f).

As stated before, a pure state of the quantum system, consisting of n qubits, is a vector of unit-length in the 2nd dimensional Hilbert space ($|\psi\rangle = \sum_{i=0}^{2^n-1} c_i |i\rangle$). Where $\{|i\rangle\}$ is a set of orthonormal basis to each other for any basis of the Hilbert space, and $c_i$ corresponds to the probability amplitudes where $\sum_i |c_i|^2 = 1$. For instance, if the quantum system consists of a single qubit, then:

$$|\psi\rangle = \sum_{i=0}^{2^n-1} c_i |i\rangle \bigg|_{n=1} \quad \rightarrow \quad |\psi\rangle = c_0|0\rangle + c_1|1\rangle \quad c_1, c_0 \in \mathbb{C}$$

The variables $c_0$ and $c_1$ are complex numbers representing probability amplitudes, which means that $|c_0|^2$ is the probability of getting $|\psi\rangle = |0\rangle$ because of the measurement on qubit $|\psi\rangle$ and $|c_1|^2$ is the probability of getting $|\psi\rangle = |1\rangle$ because of the measurement. If $c_0 = c_1$, it indicates that the state $|\psi\rangle$ is in a superposition of two states $|0\rangle$ and $|1\rangle$ with equally weighted probabilities (i.e., the probability amplitude is equally weighted for the ẑ-basis). The interpretation of values of a wave function as the probability amplitude is a pillar of the Copenhagen interpretation [8][18]. Suppose the observable Q is uncertain and in coherent superposition of the observable eigenstates when a measurement of Q is made. In that case, the system jumps to one of the eigenstates returning the eigenvalue belonging to that eigenstate.

If we use the notation of Figure 7 in order to elaborate on the provenance of $|\psi\rangle$ formulation:

$$|\psi\rangle = \alpha|0\rangle + \beta|1\rangle, \quad \alpha, \beta \in \mathbb{C} \quad (3.a)$$

$$|\psi\rangle = r_\alpha \cdot e^{j\emptyset_\alpha}|0\rangle + r_\beta \cdot e^{j\emptyset_\beta}|1\rangle \quad (3.b)$$

If we want the compaction of the global phase of a quantum state that has no observable effects ($\emptyset = \emptyset_\beta - \emptyset_\alpha$):

$$|\psi\rangle = r_\alpha|0\rangle + r_\beta \cdot e^{j\emptyset}|1\rangle \quad (3.c)$$

That is to say:

$$|\psi\rangle = z|0\rangle + (x + jy)|1\rangle \quad x, y, z \in \mathbb{R} \quad (3.d)$$

$$|\psi\rangle = \cos\theta|0\rangle + \sin\theta \cdot e^{j\emptyset}|1\rangle \text{ for } \theta \in [0, 2\pi) \text{ and } \emptyset \in [0, 2\pi) \quad (3.e)$$

Finally, we can obtain the formula of Figure 7:

$$|\psi\rangle = \cos\left(\frac{\theta}{2}\right)|0\rangle + e^{i\emptyset} \cdot \sin\left(\frac{\theta}{2}\right)|1\rangle \text{ for } \theta \in [0, \pi) \text{ and } \emptyset \in [0, 2\pi) \quad (3.f)$$

As stated before, quantum states could be either mixed or pure. Pure states give deterministic results when measured on appropriate bases, whereas mixed states give probabilistic results on all bases. A pure state is obtained when the qubit has no interaction with the outside world. Otherwise, a state is mixed when the environment affects the quantum system and becomes entangled [7][8]. Mixed states require a density matrix ($\rho$) to describe the statistical state of a quantum system. The need for a statistical description via density matrices arises because it is impossible to describe a quantum system using exclusively states represented by ket vectors (4). A typical situation in which a density matrix (or density operator) is needed includes a non-equilibrium time-evolution (e.g., due to the decoherence effect) that starts from a mixed equilibrium state. A pure state can be verified if $Tr(\rho) = Tr(\rho^2) = 1$.

$$|\psi\rangle = \alpha|00\rangle + \beta|01\rangle + \gamma|10\rangle + \lambda|11\rangle = \begin{pmatrix} \alpha \\ \beta \\ \gamma \\ \lambda \end{pmatrix} \quad (4)$$

Using eigenvalues, the spectral decomposition of density matrix $\rho$ can be expressed by using its orthonormal vectors $|\psi_i\rangle$.

$$\rho = |\psi\rangle\langle\psi| = \sum_i p_i \cdot |\psi_i\rangle\langle\psi_i| = \begin{pmatrix} \alpha^2 & 0 & 0 & 0 \\ 0 & \beta^2 & 0 & 0 \\ 0 & 0 & \gamma^2 & 0 \\ 0 & 0 & 0 & \lambda^2 \end{pmatrix}$$

The density matrix is conformed like this by using the outer product (the off-diagonal elements are quantum coherences and can be complex, and the diagonal elements must be real):

$$\rho = |\vartheta\rangle\langle\vartheta| = \begin{pmatrix} \alpha \\ \beta \end{pmatrix}(\alpha^* \quad \beta^*) = \begin{pmatrix} \alpha\alpha^* & \alpha\beta^* \\ \alpha^*\beta & \beta\beta^* \end{pmatrix} = \begin{pmatrix} |\alpha|^2 & \alpha\beta^* \\ \alpha^*\beta & |\beta|^2 \end{pmatrix}$$

Given a set of probabilities per state, we can take the sum of each weighted density matrix to construct the entire system's density matrix. For example, the following density matrix represents classical dependent probabilities:

$$\rho = p \cdot |0\rangle\langle 0| + (1-p) \cdot |1\rangle\langle 1| = \begin{pmatrix} p & 0 \\ 0 & (1-p) \end{pmatrix}$$

If real coordinates $(x, y, z \in \mathbb{R})$ of the Bloch sphere are used and taking into account that $(x, y, z) = (2 \cdot \mathrm{Re}(\rho_{01}), 2 \cdot \mathrm{Im}(\rho_{10}), \rho_{00} - \rho_{11})$ for $\mathbb{R}^3$ coordinates, we can figure up Table III based on [7]:

$$\rho = \begin{pmatrix} \rho_{00} & \rho_{01} \\ \rho_{10} & \rho_{11} \end{pmatrix} = \tfrac{1}{2} \cdot \begin{pmatrix} 1+z & x-jy \\ x+jy & 1-z \end{pmatrix} = \tfrac{1}{2} \cdot \left( \begin{pmatrix} 1 & 0 \\ 0 & 1 \end{pmatrix} + \begin{pmatrix} 0 & x \\ x & 0 \end{pmatrix} + \begin{pmatrix} 0 & -jy \\ jy & 0 \end{pmatrix} + \begin{pmatrix} z & 0 \\ 0 & -z \end{pmatrix} \right) \quad (5)$$

Table III. Relationship among the three ways of representing information (see Figure 7)

| $\mathbb{R}^3$ axis | Quantum bases | Coordinates | Vector | $\rho$ |
| --- | --- | --- | --- | --- |
| $\hat{x}$ | $\vert+\rangle$ | (1,0,0) | $\begin{pmatrix} 1 \\ 1 \end{pmatrix}$ | $\tfrac{1}{2} \cdot \begin{pmatrix} 1 & 1 \\ 1 & 1 \end{pmatrix}$ |
| $\hat{x}$ | $\vert-\rangle$ | (-1,0,0) | $\begin{pmatrix} 1 \\ -1 \end{pmatrix}$ | $\tfrac{1}{2} \cdot \begin{pmatrix} 1 & -1 \\ -1 & 1 \end{pmatrix}$ |
| $\hat{y}$ | $\vert\pi/2\rangle$ | (0,1,0) | $\begin{pmatrix} 1 \\ i \end{pmatrix}$ | $\tfrac{1}{2} \cdot \begin{pmatrix} 1 & -i \\ i & 1 \end{pmatrix}$ |
| $\hat{y}$ | $\vert-\pi/2\rangle$ | (0,-1,0) | $\begin{pmatrix} 1 \\ -i \end{pmatrix}$ | $\tfrac{1}{2} \cdot \begin{pmatrix} 1 & i \\ -i & 1 \end{pmatrix}$ |
| $\hat{z}$ | $\vert 0\rangle$ | (0,0,1) | $\begin{pmatrix} 1 \\ 0 \end{pmatrix}$ | $\begin{pmatrix} 1 & 0 \\ 0 & 0 \end{pmatrix}$ |
| $\hat{z}$ | $\vert 1\rangle$ | (0,0,-1) | $\begin{pmatrix} 0 \\ 1 \end{pmatrix}$ | $\begin{pmatrix} 0 & 0 \\ 0 & 1 \end{pmatrix}$ |

Like classical computers that work based on Boolean logic gates, quantum computers perform gate operations to manipulate quantum information. We can track the quantum system's state through a step-by-step computation by writing down bra-ket expressions. However, sometimes it is easier to understand the transition using a different mathematical representation: a common quantum circuit that consists of multiple gate operations. Using gate operations, it is possible to understand the quantum information teleportation from one place to another (Figures 2, 8, and 9). This is not only limited to close distances but also applies to long distances [15][16]. The reader must consider that the laws of quantum mechanics permit only unitary transformations over the Hilbert space, represented by unitary matrices. Actually, quantum algorithms manipulate the amplitude and phase of the quantum wave function of the system to build interference patterns, affecting the probability of measuring particular values. So, when we measure a qubit, its quantum state collapses to the observed state. The problem is that it is impossible to measure the qubit's state without changing it. Besides, the imperfections accumulated throughout the quantum operations strongly depend on the particular technology adopted for representing a qubit. This fact reveals one of the main problems that this quantum data trustworthiness layer must be coped with: decoherence [16][37].

When a system interacts with its environment, the off-diagonal elements decay, and the final density matrix is the diagonal one (i.e., a statistical mixture). This process is called decoherence. The decoherence model used in our proposal is based on (5) but adds the decoherence effects as quantum multiplicative imperfections [8][7][37][38]. For example, the short-time effect could be caused for the depolarizing phenomenon (bit and phase flip during teleportation), and the long-term effect could be caused for the thermalization noise that is of utmost importance in quantum memories. Decoherence is measured with the help of decoherence times (i.e., the time for which the qubits can be entangled without any loss of information). So, operations must be completed before the qubits lose information. It depends upon the technology used for qubits. For example, qubits realized with superconducting circuits exhibit 100 microseconds of decoherence time, and a much larger decoherence time has been reported with trapped ions [7][8].

If $\{\gamma_k\}$ are the decay rates for $(\hat{x}, \hat{y}, \hat{z})$-coordinates, the decoherence time evolution can be modeled as proposed in [7]:

$$x(t) = x(0) \cdot e^{-2t(\gamma_y + \gamma_z)}$$

$$y(t) = y(0) \cdot e^{-2t(\gamma_x + \gamma_z)}$$

$$z(t) = z(0) \cdot e^{-2t(\gamma_x + \gamma_y)}$$

Then, our mixed state model that was expressed in (5) is affected as in the following:

$$\rho(t) = \begin{pmatrix} \rho(t)_{00} & \rho(t)_{01} \\ \rho(t)_{10} & \rho(t)_{11} \end{pmatrix} =$$

$$= \frac{1}{2} \cdot \begin{pmatrix} 1 + z(0) \cdot e^{-2t(\gamma_x + \gamma_y)} & x(0) \cdot e^{-2t(\gamma_y + \gamma_z)} - jy(0) \cdot e^{-2t(\gamma_x + \gamma_z)} \\ x(0) \cdot e^{-2t(\gamma_y + \gamma_z)} + jy(0) \cdot e^{-2t(\gamma_x + \gamma_z)} & 1 - z(0) \cdot e^{-2t(\gamma_x + \gamma_y)} \end{pmatrix}$$

Actually, the term decoherence is often used to refer to any process that affects the qubits, including perturbations and imperfections, by describing how entangling interactions with the environment influence future measurements. Decoherence is a pure quantum effect to be distinguished from classical noise, as quantum imperfections are multiplicative rather than additive and, furthermore, could exhibit an asymmetric behavior [7][16][37]. The corruption of the quantum data by decoherence implies the irreversible loss of information. The imperfection of mixed states due to decoherence can be quantified by a fundamental figure of merit known as quantum fidelity (F) of a quantum state $|\psi\rangle$ [7].

$$F = \langle\psi|\rho|\psi\rangle$$

The fidelity is 1 for a pure state, and it decreases as the decoherence degrades the quality of the state. So, with F = 1, the actual state is identical to the desired state. The reliability of a teleportation system is measured with quantum fidelity, and the higher the imperfections introduced by decoherence, the lower the fidelity ($2^{-n} \leq F \leq 1$). A single qubit system has a fidelity of 50% with a completely mixed state. Similarly, in an n-qubit system, a completely mixed state has a fidelity of $F = 1/2^n$. For this reason, Quantum Error Correction (QEC) techniques are extensively adopted to preserve the quantum information against decoherence and imperfections (Figure 3). Usually, the problem of decoherence is overcome by entanglement distillation or purification, which requires an additional level of qubit processing. Therefore, QEC codes are required, and network functionalities must be designed accordingly, making the system robust. Such design is present in Figure 3, starting from the functionality layer of error-correcting to the layer of medium access control and route discovery, to the transport layer protocols.

If the mitigation of those imperfections must be obtained by the quantum data trustworthiness layer, some redundancy in the quantum information sequence (i.e., quantum error correcting codes and fault-tolerant techniques) or additional processing (i.e., purification and distillation) is imposed (see Figure 5). Worthless to say that noise is one of the main obstacles in entanglement distribution as it will degrade the entanglement. Entanglement purification distills the high-quality entangled states from the low-quality ones with the local operation and classical communications [39]. In traditional entanglement purification protocols for physical-qubit entanglement, two kinds of errors should be purified: The bit-flip error and the phase-flip error. Using two or more less-entangled mixed pairs shared among nodes, creating one pair with a higher entanglement is possible. As stated before, the introduced overhead must be taken into account in the routing metric when the shortest path is searched for, although quantum error correction can undo part of the degradation of the superposition state induced by decoherence.

### B. Quantum network trustworthiness layer

The simplest circuit implementation for quantum teleportation is shown in Figures 8 and 9. In those circuits, each measurement comes with a classical feedforward operation to the residual qubit, which is essential for completing the teleportation of an arbitrary quantum state [40]. Even though entangled particles always share physical properties regardless of the distance between them, the necessity of classical communication forbids the transmission of information from one place to another faster than the speed of light. In fact, the scheme presented in Figure 2 considers two remote parties connected by a quantum channel, which may exploit two-way classical communication and adaptive local operations. In this use case scenario, we determine the maximum trustworthiness that highly depends on achievable rates, distributing entanglement, and generating quantum consensus through the simulated quantum channels [41].

Quantum entanglement can achieve tasks that are difficult to coordinate in classical networks, and a well-known one is quantum teleportation. For that purpose, a pair of particles indicated as a Bell Pair (also known as EPR Pair or Einstein, Podolsky, and Rosen Pair) is generated and distributed to the source and the destination (see the sender scheme in Figure 2). In order to establish long-distance quantum communication, it is necessary to distribute the entangled qubits beforehand. Using gate operations, it is possible to teleport quantum information from one place to another. Our modeled circuit for quantum teleportation is shown in Figure 2. We can track the quantum global state $|\varphi_x\rangle$ through a step-by-step computation by writing down the bra-ket expressions as the following examples illustrate:

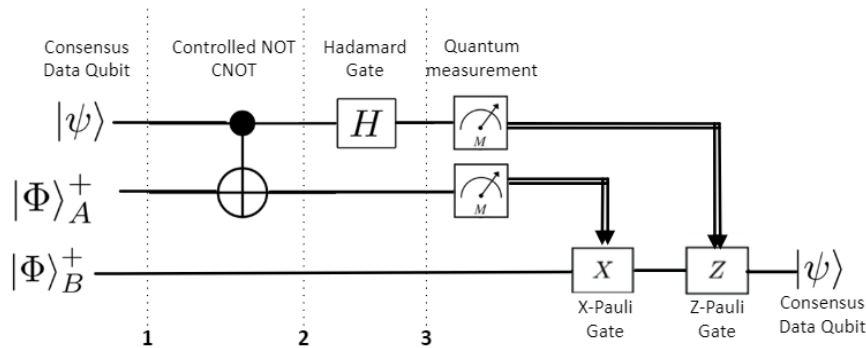

**Figure 8.** Example of teleportation by using the EPR pair $|\Phi^+\rangle$.

The first step is to recognize the global quantum state:

$$\begin{cases} |\psi_{Data}\rangle = \alpha|0\rangle + \beta|1\rangle \\ |\Phi^+\rangle = \dfrac{|00\rangle + |11\rangle}{\sqrt{2}} \end{cases}$$

$$|\varphi_x\rangle = |\psi_{Data}\rangle \otimes |\Phi^+\rangle$$

So, in track point 1 of Figure 8, we obtain the following:

$$|\varphi_1\rangle = \alpha|0\rangle \otimes \frac{|00\rangle + |11\rangle}{\sqrt{2}} + \beta|1\rangle \otimes \frac{|00\rangle + |11\rangle}{\sqrt{2}}$$

$$|\varphi_1\rangle = \frac{1}{\sqrt{2}} \cdot (\alpha|000\rangle + \alpha|011\rangle + \beta|100\rangle + \beta|111\rangle)$$

Where each qubit must be read into $|qubit_1\ origin, qubit_2\ origin, qubit_1\ destination\rangle$. In track point 2 of Figure 8, we have to apply the CNOT operator at the 1st and 2nd qubits where,

$$CNOT\ operator \triangleq \begin{pmatrix} 1 & 0 & 0 & 0 \\ 0 & 1 & 0 & 0 \\ 0 & 0 & 0 & 1 \\ 0 & 0 & 1 & 0 \end{pmatrix}$$

$$|\varphi_2\rangle = \frac{1}{\sqrt{2}} \cdot (\alpha|000\rangle + \alpha|011\rangle + \beta|110\rangle + \beta|101\rangle)$$

Hadamard operator has to be applied at the 1st qubit in track point 3 of Figure 8, where,

$$H\ operator \triangleq \frac{1}{\sqrt{2}} \begin{pmatrix} 1 & 1 \\ 1 & -1 \end{pmatrix} \text{ and thus, } |0\rangle \equiv \frac{|0\rangle + |1\rangle}{\sqrt{2}} \text{ and } |1\rangle \equiv \frac{|0\rangle - |1\rangle}{\sqrt{2}}$$

$$|\varphi_3\rangle = \frac{1}{2} \cdot (\alpha|000\rangle + \alpha|100\rangle + \alpha|011\rangle + \alpha|111\rangle + \beta|010\rangle - \beta|110 + \beta|001\rangle - \beta|101\rangle)$$

$$|\varphi_3\rangle = \frac{1}{2} \cdot [|00\rangle \otimes (\alpha|0\rangle + \beta|1\rangle) + |01\rangle \otimes (\alpha|1\rangle + \beta|0\rangle) + |10\rangle \otimes (\alpha|0\rangle - \beta|1\rangle) + |11\rangle \otimes (\alpha|1\rangle - \beta|0\rangle)]$$

Finally, the global quantum state has four possible quantum states with a 25% probability:

$$\text{If } |00\rangle \rightarrow \begin{pmatrix} \alpha \\ \beta \end{pmatrix} \text{ then } |\psi_{Data}\rangle$$

Else, a transformation is needed by using $X \equiv \begin{pmatrix} 0 & 1 \\ 1 & 0 \end{pmatrix}$ and $Z \equiv \begin{pmatrix} 1 & 0 \\ 0 & -1 \end{pmatrix}$

$$\text{If } |01\rangle \rightarrow \begin{pmatrix} \beta \\ \alpha \end{pmatrix} \text{ then } X \begin{pmatrix} \beta \\ \alpha \end{pmatrix} = \begin{pmatrix} \alpha \\ \beta \end{pmatrix} = |\psi_{Data}\rangle$$

$$\text{If } |10\rangle \rightarrow \begin{pmatrix} \alpha \\ -\beta \end{pmatrix} \text{ then } Z \begin{pmatrix} \alpha \\ -\beta \end{pmatrix} = \begin{pmatrix} \alpha \\ \beta \end{pmatrix} = |\psi_{Data}\rangle$$

$$\text{If } |11\rangle \rightarrow \begin{pmatrix} -\beta \\ \alpha \end{pmatrix} \text{ then } XZ \begin{pmatrix} -\beta \\ \alpha \end{pmatrix} = \begin{pmatrix} \alpha \\ \beta \end{pmatrix} = |\psi_{Data}\rangle$$

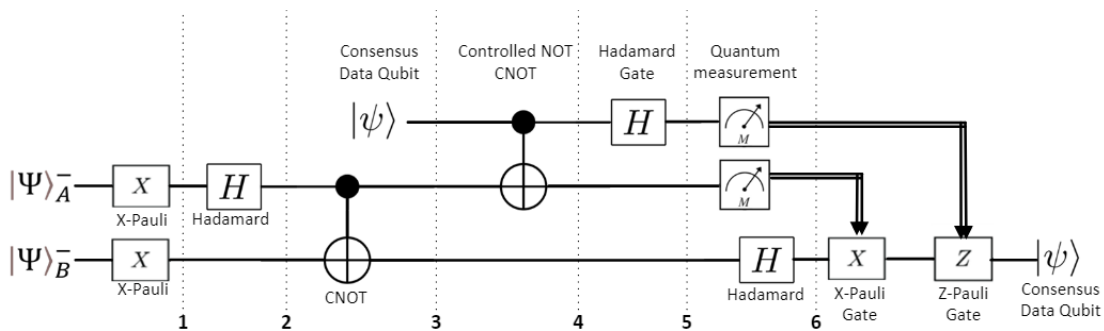

**Figure 9.** Example of teleportation by using the EPR pair $|\Psi^-\rangle$.

The first step for the second example (**Figure 9**) is also to recognize the global quantum state:

$$\begin{cases} |\psi_{Data}\rangle = \alpha|0\rangle + \beta|1\rangle \\ |\Psi^-\rangle = \dfrac{|01\rangle - |10\rangle}{\sqrt{2}} \end{cases}$$

$$|\varphi_x\rangle = |\psi_{Data}\rangle \otimes |\Psi^-\rangle$$

$$|\varphi_1\rangle = \alpha|0\rangle \otimes \frac{|01\rangle - |10\rangle}{\sqrt{2}} + \beta|1\rangle \otimes \frac{|01\rangle - |10\rangle}{\sqrt{2}}$$

$$|\varphi_1\rangle = \frac{1}{\sqrt{2}} \cdot (\alpha|001\rangle - \alpha|010\rangle + \beta|101\rangle - \beta|110\rangle)$$

So, in track point 1 of Figure 9, we have obtained the following global state after the X operator is applied at the 2nd and 3rd qubits:

$$|\varphi_1\rangle = \frac{1}{\sqrt{2}} \cdot (\alpha|010\rangle - \alpha|001\rangle + \beta|110\rangle - \beta|101\rangle)$$

Hadamard operator at the 2nd qubit has been applied in track point 2 of Figure 9:

$$|\varphi_2\rangle = \frac{1}{2} \cdot (\alpha|000\rangle - \alpha|010\rangle - \alpha|001\rangle - \alpha|011\rangle + \beta|100\rangle - \beta|110\rangle - \beta|101\rangle - \beta|111\rangle)$$

Then, CNOT operator is applied at the 2nd and 3rd qubits in track point 3 of Figure 9:

$$|\varphi_3\rangle = \frac{1}{2} \cdot (\alpha|000\rangle - \alpha|011\rangle - \alpha|001\rangle - \alpha|010\rangle + \beta|100\rangle - \beta|111\rangle - \beta|101\rangle - \beta|110\rangle)$$

After the application of the CNOT operator at the 1st and 2nd qubits:

$$|\varphi_4\rangle = \frac{1}{2} \cdot (\alpha|000\rangle - \alpha|011\rangle - \alpha|001\rangle - \alpha|010\rangle + \beta|110\rangle - \beta|101\rangle - \beta|111\rangle - \beta|100\rangle)$$

Hadamard operator at the 1st qubit has been applied in track point 5 of Figure 9:

$$|\varphi_5\rangle = \frac{1}{2 \cdot \sqrt{2}} \cdot (\alpha|000\rangle + \alpha|100\rangle - \alpha|011\rangle - \alpha|111\rangle - \alpha|001\rangle - \alpha|101\rangle - \alpha|010\rangle - \alpha|110\rangle + \beta|010\rangle$$
$$- \beta|110\rangle - \beta|001\rangle + \beta|101\rangle - \beta|011\rangle + \beta|111\rangle - \beta|000\rangle + \beta|100\rangle)$$

$$|\varphi_5\rangle = \frac{1}{2 \cdot \sqrt{2}} \cdot [|00\rangle \otimes (\alpha|0\rangle - \alpha|1\rangle - \beta|1\rangle - \beta|0\rangle) + |01\rangle \otimes (-\alpha|1\rangle - \alpha|0\rangle + \beta|0\rangle - \beta|1\rangle) + |10\rangle$$
$$\otimes (\alpha|0\rangle - \alpha|1\rangle + \beta|1\rangle + \beta|0\rangle) + |11\rangle \otimes (-\alpha|1\rangle - \alpha|0\rangle - \beta|0\rangle + \beta|1\rangle)]$$

$$|\varphi_5\rangle = \frac{1}{2 \cdot \sqrt{2}} \cdot [|00\rangle \otimes \binom{(\alpha - \beta)}{(-\alpha - \beta)} + |01\rangle \otimes \binom{(-\alpha + \beta)}{(-\alpha - \beta)} + |10\rangle \otimes \binom{(\alpha - \beta)}{(-\alpha + \beta)} + |11\rangle \otimes \binom{(-\alpha - \beta)}{(-\alpha + \beta)}]$$

Hadamard operator at the 3rd qubit has been finally applied in track point 6 of Figure 9:

$$|\varphi_6\rangle = \frac{1}{2} \cdot [|00\rangle \otimes \binom{-\beta}{\alpha} + |01\rangle \otimes \binom{-\alpha}{\beta} + |10\rangle \otimes \binom{\beta}{\alpha} + |11\rangle \otimes \binom{-\alpha}{-\beta}]$$

Finally, the global quantum state also has four possible quantum states with a 25% probability:

$$\text{If } |11\rangle \rightarrow \binom{-\alpha}{-\beta} \text{ then } |\psi_{Data}\rangle \text{ with a global phase of } \pi,$$

Else, a transformation is needed via X and Z Pauli's matrices:

$$\text{If } |10\rangle \rightarrow \binom{\beta}{\alpha} \text{ then } X\binom{\beta}{\alpha} = \binom{\alpha}{\beta} = |\psi_{Data}\rangle$$

$$\text{If } |01\rangle \rightarrow \binom{-\alpha}{\beta} \text{ then } Z\binom{-\alpha}{\beta} = \binom{-\alpha}{-\beta} = -|\psi_{Data}\rangle$$

$$\text{If } |00\rangle \rightarrow \binom{-\beta}{\alpha} \text{ then } XZ\binom{-\beta}{\alpha} = \binom{\alpha}{\beta} = |\psi_{Data}\rangle$$

The quantum teleportation technique introduced above is an example of transmitting one bit of quantum information from one place to another. However, networking often requires multi-hop communication. The simplest solution for such demand is to apply quantum teleportation hop-by-hop. Nevertheless, operating on the message qubit directly many times degrades the information. Another solution is the use of entanglement swapping [42][43][44], which is capable of lengthening the Bell pair to allow direct teleportation of quantum information over multiple repeaters.

Figure 5 introduces quantum data purification/distillation and entanglement purification at the quantum network trustworthiness layer. Entanglement purification is a technique that could increase the fidelity of entangled pairs by using two or more less-entangled mixed pairs ($F_{input}$) and by creating one pair with a higher entanglement ($F_{output}$) (6) [39]. As it is said, the reliability of a teleportation system is measured with quantum fidelity. So, the imperfection of a quantum state can be described by its fidelity (e.g., with 100% of fidelity, the actual state is identical to the desired pure state, and with 50% of fidelity, a qubit is in a completely mixed state). The larger the imperfections introduced by decoherence, the lower the fidelity. For high-fidelity end-to-end, several purification and swapping iterations have to be executed (Figure 10).

$$F_{output} = \frac{F_{input}^2}{F_{input}^2 + (1 - F_{input})^2} \quad (6)$$

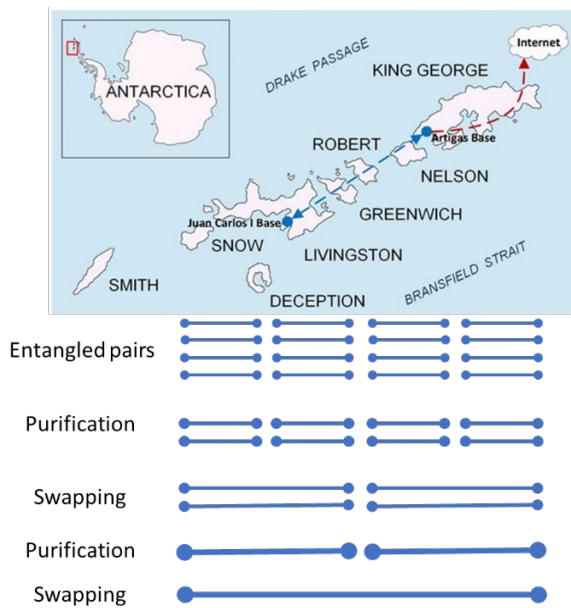

**Figure 10.** Example of entanglement swapping and purification.

The evaluation of the trustworthiness of a single node exclusively based on its forwarding behavior may result in an inaccurate trust computation of a node's trust value, so the network reliability is also a paramount indicator of the system's trustworthiness. Holistic reliability is needed for the upper layer protocol of the quantum Internet architecture (Figure 3): the application layer that directly provides services for permafrost telemetry. A secure and distributed strategy to compute global trust values must be deployed in simulated scenarios. Although the model of Figure 5 makes explicit cyber-security issues regarding intentional malicious digital attacks, malicious nodes modeling is out of the scope of our trustworthiness modeling as these nodes are beyond the definition of Antarctica's environment in the telemetry use case of SHETLAND-NET research project [14][45].

We acknowledge that this fact leaves aside from our studies one of the most well-known strengths of quantum Internet (i.e., when cryptography exploits the mechanical principle of non-cloning). Well-known quantum protocols such as Quantum Key Distribution (QKD), which generates secure keys between separated parties, make proper use of an inherently secure quantum channel because any attempt of hacking will disturb the state of the shared key qubits, thus exposing this malicious attempt [9][15][46]. Provable attacks against unlimited eavesdropping attacks are provided in this way. Actually, QKD does not simply rely on the difficulty of solving a mathematical problem. Therefore, even a quantum computer could not break the key. The origin of the security of QKD can be traced back to the fundamental quantum physical principles of superposition and no-cloning. Another example is the Quantum Secure Direct Communication (QSDC), which transmits private information directly between communication parties without producing secret keys in advance [10].

Despite this, we believe that the quantum Internet architecture should enable to use the consensus layer minimizing traffic congestion and avoiding worsening reliability due to the link saturation effect. Our research wants to take advantage of the multi-objective problem approach by exploiting the complexity reduction offered by quantum parallelism and exploiting the instantaneous coordination of the communication parties thanks to quantum teleportation and the superposition of qubit states [47]. Several aspects are relevant in determining the quality of a Byzantine agreement protocol. As for most protocols, the amount of local computation and the total number of message bits exchanged continue to be important. Nevertheless, in this archetypal problem in distributed adversarial computation, two are the most relevant (and most investigated) aspects: the round complexity and the fault model [15][49]. In [48][49], authors can reach a general agreement in $O(1)$ expected communication rounds tolerating up to the optimal (TOTAL_PLAYERS-1)/3 faulty players in the synchronous setting (fast quantum byzantine agreement). Besides, the quantum Internet offers much lower time and communications complexity, in terms of message number, than the classical Internet. The use of quantum Internet for the sake of trustworthiness via fast quantum consensus is the main goal of the research presented in this paper.

## 4. Assessing and quantifying trustworthiness by simulations

Several indicators can be defined to assess the trustworthiness level in a quantum Internet that includes classical Internet (see Figure 2 and Figure 3). We have defined a quantitative indicator for each layer to break down the trustworthiness model and improve the communication system's performance (see Figure 5). In this regard, we propose to use four well-known key performance indicators as they are also used to measure the impact of the technologies and approaches described in the literature [5][6]. Our simulations must help us identify our research's possible weaknesses in a quantum telemetry system that pursues the maximization of trustworthiness "at any price".

To the best of our knowledge, none of the prior analyzed trustworthiness approaches have included neither the four trustworthiness dimensions nor the consideration of the interdependencies between them [28][29][30][31][32][33][34][35]. Our simulations must help us anticipate and identify the possible weaknesses of our IoT telemetry system [11]. Our reliability simulated model includes four main indicators that have been included in all the simulations.

1. Faulty sensing ratio (FSR): It is defined as the proportion of false sensed values by all nodes and total sensed values in a defined observation period. Thus, the lower the faulty sensing ratio is, the better the data trustworthiness. Corrective methods can be applied to improve this indicator (e.g., hashes, parity bits, and checksums) and compensate for sensor malfunctioning or sensed data misreading. In our simulations, it is supposed that no corrective methods are used in the system. In that case, sensed data (e.g., frozen ground status) are considered to be faultily sensed if the value stored in the sensing node is different from the value that the sensor should have correctly read.

2. Packet delivery ratio (PDR): It is defined as the relationship between the total number of packets correctly received and the total number of sent packets by all nodes in the same time slot. Obviously, the higher the packet delivery ratio is, the better the network's trustworthiness. Delay tolerant techniques, network robustness, and quality of service awareness are helpful mechanisms to increase this indicator by balancing it against main issues such as link failures and network congestion.

3. Successful transaction rate (STR): It is defined as the relationship between the number of successful transactions and the total number of transactions in an observed time slot. A transaction is considered correctly achieved when a node obtains some information from a source node before a defined maximum reception time. It is a valuable indicator to compute the node reputation (i.e., social trustworthiness) based on previous transactions to build a notion of the node's reliability. In our simulations, we pursued the maximization of this key performance indicator as the main metric for the trustworthiness of the deployed architecture. Our simulations showed that the STR is highly correlated with the other figures of merit since it aggregates most of the quantitative information provided by the other indicators (Figure 5).

4. Byzantine node tolerance (BNT): It is defined as the proportion of supported byzantine nodes that can participate in the consensus system without affecting the correctness of the general agreement. A byzantine node could suffer from a crash or a soft fault by producing a malfunction. The higher the byzantine node tolerance is, the higher the probability of achieving a trustworthy general agreement. Theoretically, if the number of byzantine nodes is higher than 50% of the total number of participating nodes, none of the consensus mechanisms will reach a benevolent agreement. It is worth considering that, typically, the algorithms required to support byzantine behaviors require exchanging many messages between nodes to reach a consensus, which shall degrade the overall network performance. The experimentation performed in this paper tries to minimize this issue by using quantum Internet.

The primary goal of simulations of the SHETLAND-NET's use case is to increase the successful transaction rate (STR) to provide better reliability and achieve the needed performance (i.e., at least 60% of successful transactions) for permafrost telemetry [1][4]. Note that each proposed key performance indicator can be improved by implementing countermeasures by assuming the, sometimes, unexpected trade-offs (e.g., computational, economic, and throughput costs) [11]. For instance, although the classical consensus mechanism mitigates byzantine errors thanks to the general agreement mechanism, it negatively affects the packet delivery ratio due to the considerable amount of extra traffic injected into the network, which could cause an increment of congestion traffic unless a quantum consensus management plane is implemented (see Figure 3). This is, at least, what we postulate in this paper and the main subject for our research experiment.

## 5. Experimental evaluation and final discussion

The permafrost telemetry scenario has been represented and evaluated in the Riverbed Modeler simulator [3]. To perform the simulation tests and assess the results using our proposed trustworthiness model, we have modeled the classical communication media (the physical and link layers of LoRa and NVIS technologies), the link layer mechanisms of the quantum Internet (quantum pre-processing, quantum post-processing and entanglement generation/distribution), the telemetry application, the faulty behavior of byzantine nodes, the social trust management and the consensus algorithms with the quantum management plane, and the DTN mechanism. A DTN has to be implemented because NVIS availability varies between 70% and 100% depending on the ionosphere state, which is highly correlated to solar activity.

The simulation scenario has five NVIS concentrator nodes, each providing an independent LoRa coverage area with its own sensors for telemetry (see Figure 2). Redundant sensors measuring the same data have been simulated at each measuring spot to assess the goodness of social and consensus layers. We have defined up to five additional redundant sensor nodes deployed in each measuring spot so that one byzantine node could be tolerated. According to our Antarctica use case experiment, we have considered a simulation scenario of 32 and 64 permafrost measuring spots for the evaluation of the goodness of quantum Internet novelties [5][6][11].

We have obtained the STR during simulation time for the whole simulation duration of 400 days to analyze the system's trustworthiness between two consecutive Antarctic campaigns (once per year in southern summer). Also, various values for byzantine fault probability ($Pb_0$) ranging from $10^{-1}$ to $10^{-3}$ have been simulated to assess the effects of using different sensor sources and battery charge levels. There are four different operative scenarios depending on which redundant-related mechanisms are used: the "Standard" mode, which does not use the social nor the consensus mechanism; the "Social" mode, the "Consensus" mode; and the "Social + Consensus" mode. These simulation tests have been run multiple times to ensure statistically significant results.

In Figure 11 (classical Internet), the number of redundant sensors per cluster (from 1 to 5) and the number of measuring spots (32 and 64) vary on the horizontal axis. Therefore, the "Measuring Spots x Redundant Sensors" axis has ten discrete points. We have the "Byzantine Fault Probability" in the depth axis, which has nine discrete points corresponding to the nine different $P_{b0}$ values (equally distributed between $10^{-1}$ to $10^{-3}$). Finally, we obtained STR after the simulation on the vertical axis. A colored surface represents simulation results for each of the four trustworthiness modes.

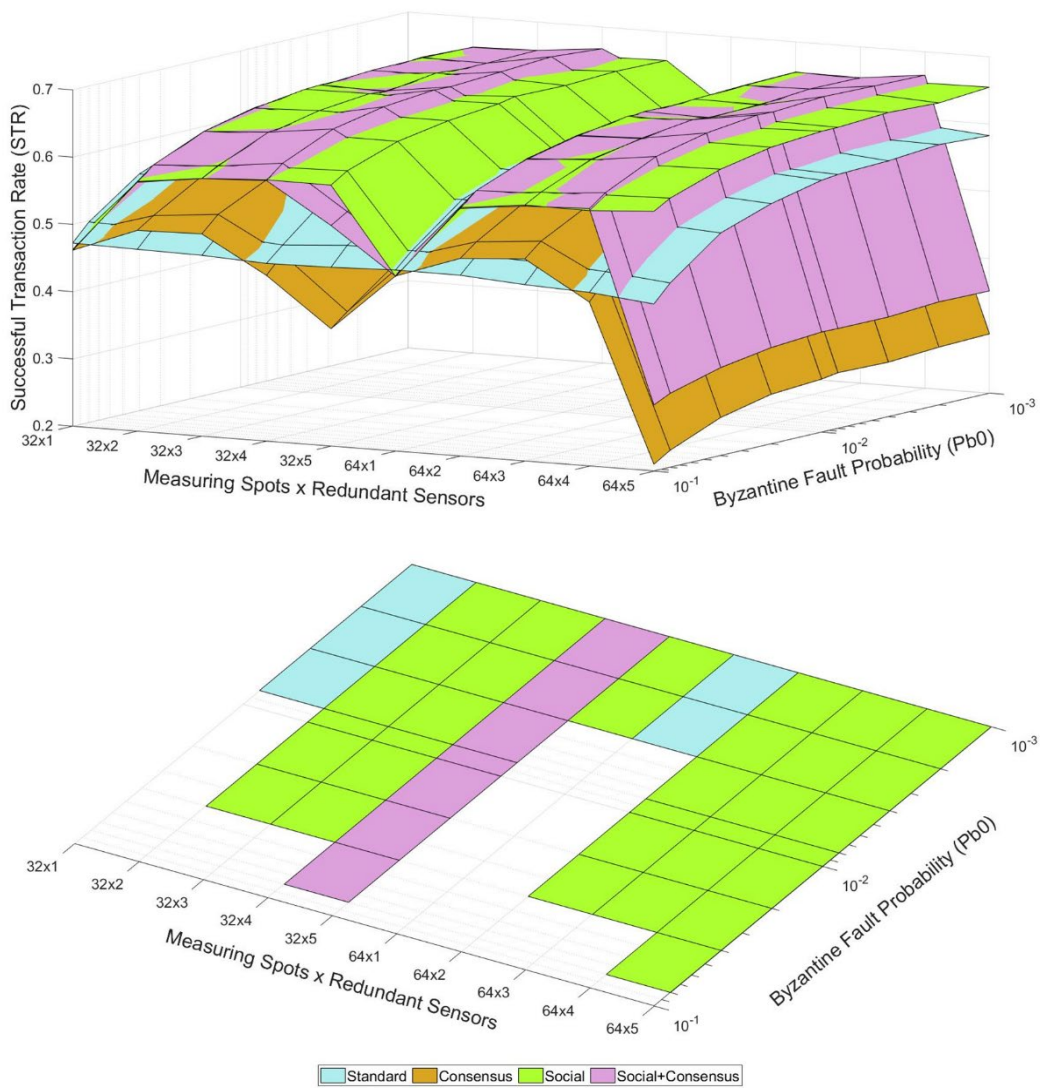

**Figure 11.** Colored mesh and "STR = 0.6" slice for STR analysis (classical Internet).

Figure 11 is also used to visualize the projection of the working domain in which to implement the permafrost telemetry service by fixing the desired minimum trustworthiness level (somehow as a slice representing the plant of the elevation pattern). As stated before, our use case requires a minimum annual STR of 0.6 to meet the objective of researchers' quality of experience [4][6]. For every point in the grid, if no solution provides an STR higher than the desired minimum value, the surface for that area is white-colored, meaning that we cannot deploy the service under these conditions. In general, the surface is painted with the color of the solution with the highest STR (Figure 11 for classical Internet and Figure 12 for quantum Internet).

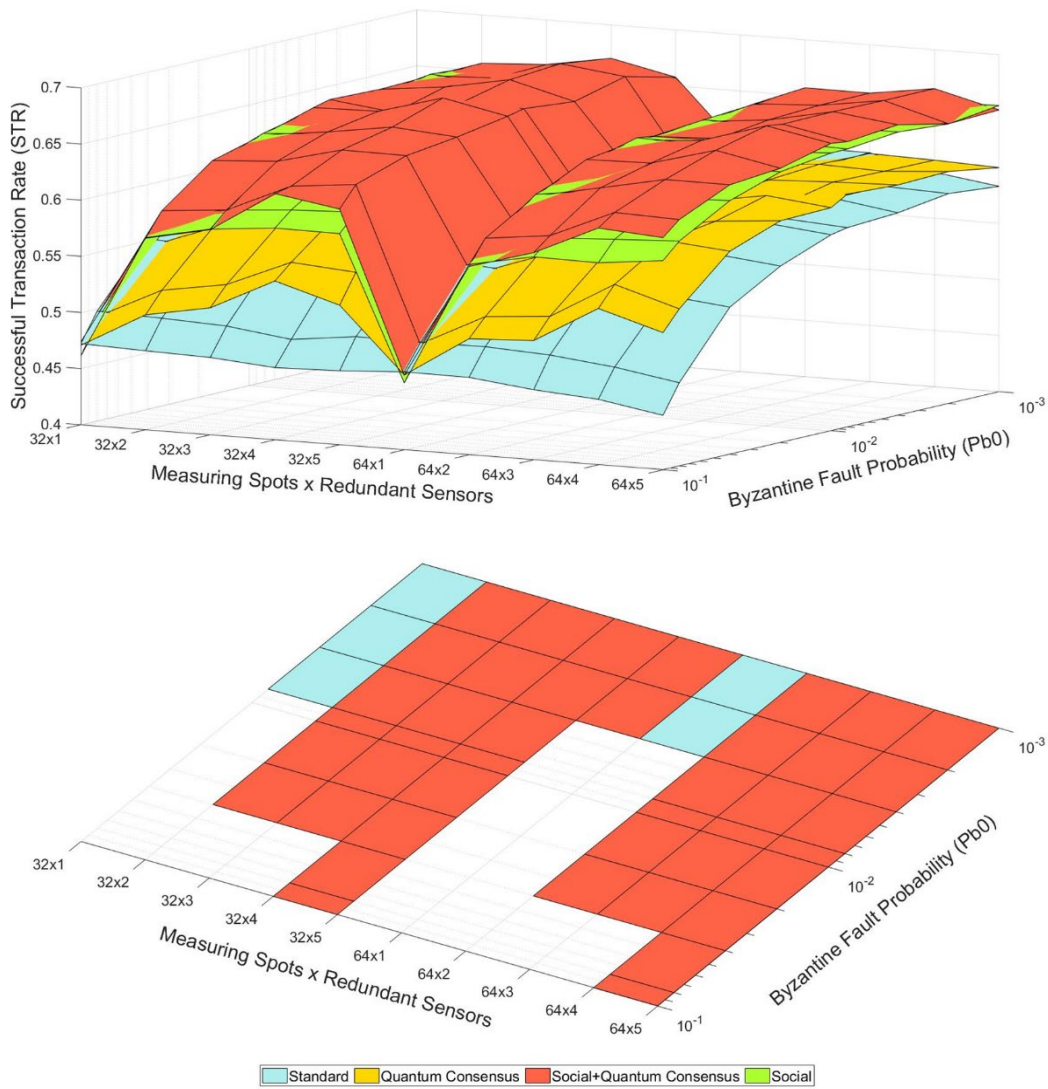

**Figure 12.** Colored mesh and "STR = 0.6" slice for STR analysis (quantum Internet).

Several conclusions can be extracted from Figures 11 and 12 regarding the benefits in trustworthiness when fast quantum consensus [48][49] is used instead of classical consensus. They are described in the following:

1. Across the "Measuring Spots x Redundant Sensors" axis, we can observe a general decrease as the number of sensors increases. This behavior is expected because the LoRa access network has a low bitrate capacity (5.47 kbps), so the network can be congested even if no consensus protocol is used. As expected, we can also observe a general decrease as the byzantine fault probability increases.

2. In Figure 11, although the "Social + Consensus" mode usually decreases below 0.6 when congestion traffic increases since the general agreement protocol overwhelms the underlying data network, we can observe an interesting zone (purple zone) where consensus positively contributes to STR performance when four redundant sensor nodes are deployed, by producing a tolerance to one byzantine node in every measuring spot, which is the main advantage of the consensus mechanism. Out of this purple zone, the "Social" mode mainly surpasses the rest of the modes. The "Social" mode seems to be the most robust one, consistently achieving the trustworthiness objective. In fact, the dropped packets in the "Social" mode are caused by common channel errors from NVIS and LoRa networks and not by network congestion. Nevertheless, the "Standard" mode is obviously the only suitable one if we cannot afford redundant sensors in any measuring spot.

3. Regarding social trustworthiness, the ostracism process cannot be abrupt in real-world scenarios since condemned nodes would never recover. This explains why correct nodes inherit the trust degradation of the byzantine node at the beginning while the byzantine node is still not condemned to ostracism. Similarly, on the right side of Figure 13, a small ripple can be seen since other nodes still try to contact the byzantine node to check whether it has recovered. In fact, it is unrealistic that a node has a byzantine behavior uniformly distributed over time [35]. Also, it is interesting to see how the other nodes (that are safe and correct) inherit the effects of the byzantine behavior, which degrades their trust. This happens until they all agree on pushing the byzantine node to the ostracism and, thus, ignore its messages, which enables them to recover their trust. For the sake of the experiment, Node D has been excluded from all the communications involving the byzantine node, which results in a convenient way to see the ideal trust figure for a node.

4. If we compare the classical (Figure 11) versus the quantum Internet (Figure 12), we observe how, in the first case, the best solution often is the "Social" mode because the use of the consensus algorithm worsens the achieved STR due to the large number of packets that it introduces to the network (traffic congestion). On the contrary, with quantum consensus, the best option is often the "Social + Quantum Consensus" mode, given that the benefits of the consensus algorithm (i.e., the byzantine node tolerance) are added to the social trust management protocol without increasing the network load. This fact allows us to gather more reliable data using the same deployed monitoring infrastructure.

5. The best case in the classical Internet simulations is the "Social + Consensus mode" mode with 32 measuring spots and 4 redundant sensors per spot, achieving an STR of 0.67. On the other hand, the quantum Internet architecture achieves a maximum STR of 0.69 in the "Social + Quantum Consensus" mode with the same configuration. So, quantum scenarios achieve a higher maximum STR and higher mean STR. In Table IV, we can observe how the "Social + Quantum Consensus" mode achieves the highest maximum STR and the highest mean STR, thus being more reliable, on average, than the classical Internet. "Consensus" and "Social + Consensus" modes of the classical Internet achieve the lowest average STRs of 0.52 and 0.54, respectively, while the same modes of the quantum Internet improve their average STR to 0.59 (+13%) and 0.64 (+19%), respectively.

**Table IV:** Maximum and average STR for each operation mode.

|         | Standard | Social | Consensus | Quantum Consensus | Social + Consensus | Social + Quantum Consensus |
|---------|----------|--------|-----------|-------------------|--------------------|----------------------------|
| **Max**     | 0.61     | 0.66   | 0.6       | 0.62              | 0.67               | 0.69                       |
| **Average** | 0.56     | 0.63   | 0.52      | 0.59              | 0.54               | 0.64                       |

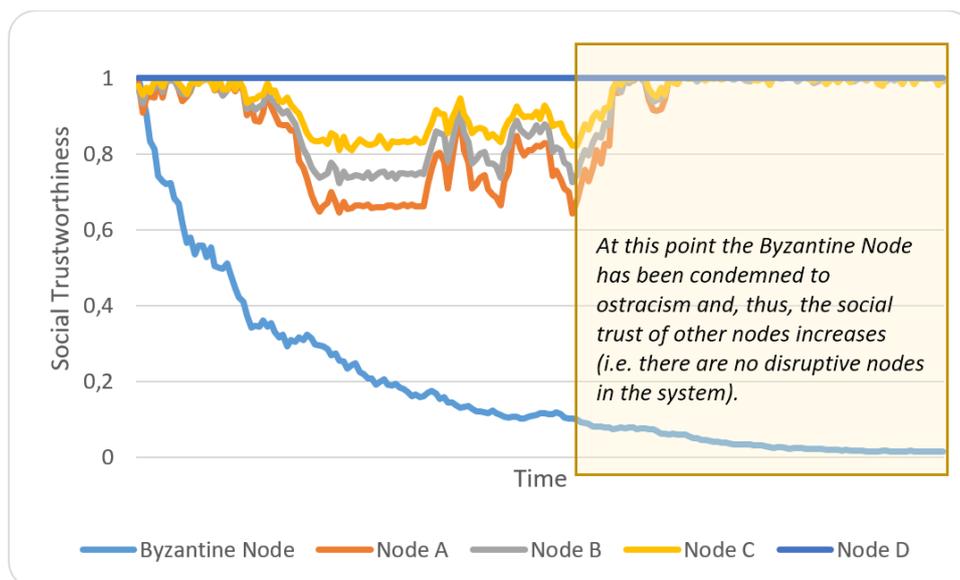

**Figure 13.** Evolution of social trustworthiness in a single measuring spot.

To sum up, in this work, we aimed to improve the trustworthiness of the Antarctic telemetry service that will be deployed in the next campaign in the field by incorporating a quantum Internet architecture. With the quantum Internet architecture, we try to improve the performance of the consensus algorithm that provides byzantine node tolerance to our system, which showed improvable results in previous studies. We performed several simulations to compare the newly designed architecture with the previous, classical one. The results show a significant improvement of the consensus algorithm reliability in terms of achieved STR (+13% and +19% improvements in average STR) and consolidate the "Social + Quantum Consensus" mode as the best operation mode for our telemetry service, improving the trustworthiness of the best solution mode of the classical network architecture. Moreover, thanks to the quantum consensus algorithm, the network can afford more nodes (sensors) since the quantum algorithm introduces minimal extra traffic compared to the classical consensus algorithm, which already congested the network with few sensors.

As stated before, these results can be generalized to other use cases. For example, in the studied permafrost telemetry use case, the number of redundant sensors is limited to 5, so only one byzantine node is tolerated per consensus group. However, we can further assess the trustworthiness improvement of quantum consensus by increasing the number of redundant sensors. Figure 14 shows the simulation results if we set the number of redundant sensors from 4 to 10, which increases the number of tolerated byzantine nodes from 1 to 3. This figure compares the results of the "Social + Quantum Consensus" and the "Social + Consensus" modes. For this reason, using less than 4 redundant sensors would not make sense, given that no byzantine nodes could be tolerated, and using a consensus mechanism would be useless. In Figure 14, we can appreciate the impact of quantum consensus versus classical consensus, which was less visible in the results of our use case.

In this generalization, we can observe that using 5 or more redundant sensors with the classical consensus mechanism degrades the obtained STR below 0.6, and it even drops to 0 with more than 7 redundant sensors and 32 measuring spots or more than 5 sensors and 64 measuring spots. This happens because of the exponential growth in complexity and the number of messages exchanged between the group members to reach a consensus, which congests the low-bandwidth network. On the contrary, the quantum consensus mechanism does not imply an exponential growth in complexity (as stated before, its cost is $O(1)$ [48][49]) and the number of exchanged messages. This way, the network can remain uncongested while using more nodes, which keeps the obtained STR to values above 0.6.

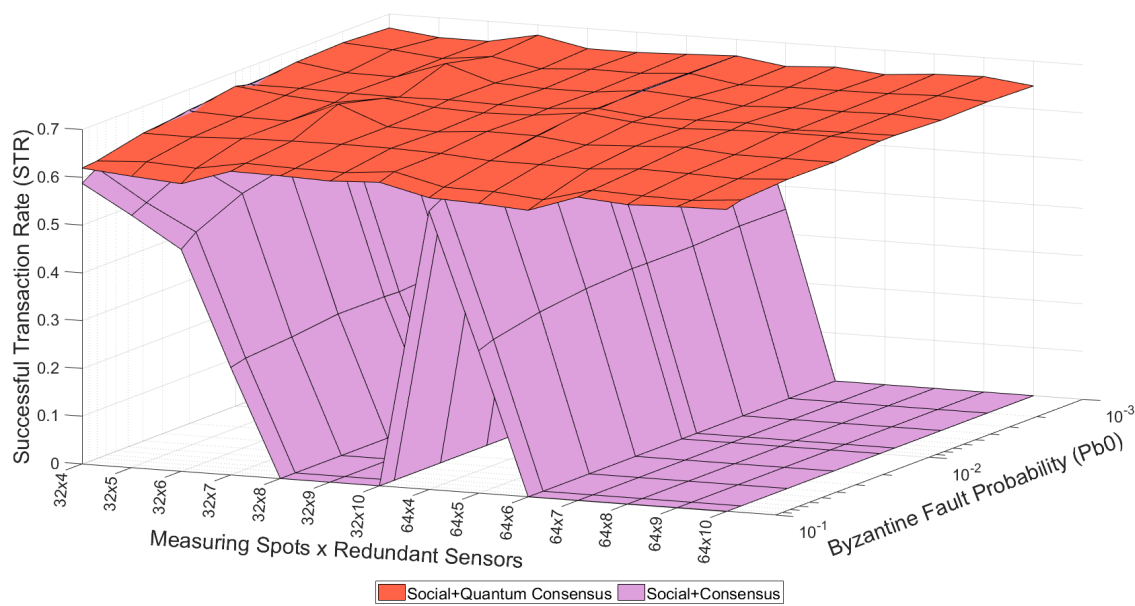

**Figure 14.** Trustworthiness mesh varying the number of redundant sensors from 4 to 10.

If we analyze the average STR results of these simulations (Table V), we can spot a big difference in the improvement of the quantum consensus. In the first round of tests (Antarctica's use case), "Quantum Consensus" and "Social + Quantum Consensus" improved their analogous classical modes by 13% a 19%, respectively. In contrast, we observe that in this second round (generalization), the "Quantum Consensus" mode improves the "Consensus" mode by 247%, and the "Social + Quantum Consensus" mode improves the "Social + Consensus mode" by 195%. These results remark that the use of the quantum consensus mechanisms gains more relevance as the number of members in the consensus group increases, given that the complexity increase in the classical consensus is exponential, which is avoided with quantum consensus, affecting the performance of the classical network. Thus, the quantum consensus enables broader scenarios with more sensors and measuring spots deployed.

**Table V:** Maximum and average STR for each operation mode in the second round of tests.

|  | Standard | Social | Consensus | Quantum Consensus | Social + Consensus | Social + Quantum Consensus |
|---|---|---|---|---|---|---|
| **Max** | 0.61 | 0.66 | 0.57 | 0.62 | 0.67 | 0.69 |
| **Average** | 0.54 | 0.64 | 0.17 | 0.59 | 0.22 | 0.65 |

In this context, several dimensions have arisen to the surface during simulations because of the existence of transversal effects, such as traffic congestion behavior derived from the NVIS IoT backbone or node ostracism used by the social trustworthiness model. The needed trade-off to balance the system reliability and the components' harmonization is not trivial.

**Acknowledgments:** This research was funded by the "Secretaria d'Universitats i Recerca del Departament d'Empresa i Coneixement de la Generalitat de Catalunya", the European Union (EU) and the European Social Fund (ESF) 2022 FI_B2 00026]. This work also received funding from the Spanish Ministry on Science, Innovation and University, the Investigation State Agency and the European Regional Development Fund (ERDF) under the grant number RTI2018-097066-B-I00 (MCIU/AEI/FEDER, UE) for the project "NVIS Sensor Network For The South Shetland Islands Archipelago" (SHETLAND-NET).

31 of 32[48] Michael Ben-Or and Avinatan Hassidim. 2005. Fast quantum byzantine agreement. In Proceedings of the thirty-seventh annual ACM symposium on Theory of computing (STOC '05). Association for Computing Machinery, New York, NY, USA, 481–485. https://doi.org/10.1145/1060590.1060662.

[49] Garay, J. (2008). Optimal Probabilistic Synchronous Byzantine Agreement. In: Kao, MY. (eds) Encyclopedia of Algorithms. Springer, Boston, MA. https://doi.org/10.1007/978-0-387-30162-4_269.

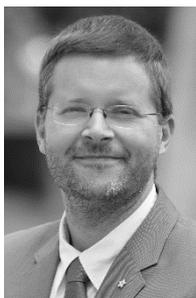 **Agustín Zaballos** was born in Madrid, Spain, in 1976. He received the M.S. and Ph.D. degrees in electronics engineering from the La Salle, Universitat Ramon Llull, in 2000 and 2012, respectively. He also holds an International MBA from La Salle URL since 2014. He is a Professor in the Computer Science Department and Project Manager of the R&D Networking Area since 2002, and is currently the head of the Research Group in Internet Technologies and Storage (GRITS). He is the Scientific Director of BitLaSalle (Barcelona Institute of Technology LaSalle), in charge of the business development and operations management of multidisciplinary research projects in La Salle Campus Barcelona (Computer Science, Internet Technologies, Telecommunications, Multimedia, Architecture, Management, and Arts&Design).

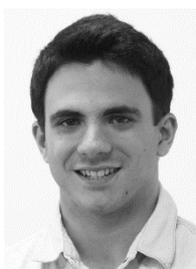 **Adrià Mallorquí** was born in Calella, Barcelona, Spain in 1995. He received the B.S. and M.S. degrees in network and telecommunication engineering from the La Salle, Universitat Ramon Llull, in 2017 and 2019, respectively. He is currently a Ph.D. candidate in the same university. From 2014 to 2017, he was a Research Intern in the Research Group on Internet Technologies and Storage (GRITS) in La Salle URL. In 2017 he became a Research Assistant at the same group and an Assistant Professor in the Engineering Department, and he started his Ph.D. in 2020.

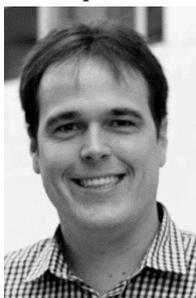 **Joan Navarro** was born in Barcelona, Spain, in 1985. He received his Ph.D. in Information and Communication Technologies from La Salle, Universitat Ramon Llull, in 2015, his MSc. degree in Telecommunications Engineering in 2008, and his BSc in Network Engineering in 2006. He is currently an Associate Professor at the Computer Engineering Department of the La Salle - URL and a member of the Group of Research in Internet Technologies and Storage of the same University. His research is focused on Cloud Computing, Big Data and Distributed Systems, specifically on the areas of Concurrency Control in Large-Scale Distributed Systems and Replication Policies in Cloud-based Databases. In addition, he is currently participating in several R&D projects.